\documentclass[twocolumn]{bmcart}

\usepackage{amsthm,amsmath}
\usepackage{amssymb}
\usepackage{pdflscape}
\usepackage[utf8]{inputenc}
\usepackage{lmodern}        
\usepackage{longtable} 
\usepackage{url} 
\usepackage{graphicx}
\usepackage{cite}

\begin{document}
\begin{frontmatter}
\begin{fmbox}

\title{BioPreDyn-bench: benchmark problems for kinetic modelling in systems biology}

\author[
   addressref={aff1},                   
]{\inits{AF}\fnm{Alejandro F} \snm{Villaverde}}
\author[
   addressref={aff1,aff2},                   
]{\inits{D}\fnm{David} \snm{Henriques}}
\author[
   addressref={aff3},                   
]{\inits{K}\fnm{Kieran} \snm{Smallbone}}
\author[
   addressref={aff4},                   
]{\inits{S}\fnm{Sophia} \snm{Bongard}}
\author[
   addressref={aff4},                   
]{\inits{J}\fnm{Joachim} \snm{Schmid}}
\author[
   addressref={aff5,aff6},                   
]{\inits{D}\fnm{Damjan} \snm{Cicin-Sain}}
\author[
   addressref={aff5,aff6},                   
]{\inits{A}\fnm{Anton} \snm{Crombach}}
\author[
   addressref={aff2},                   
]{\inits{J}\fnm{Julio} \snm{Saez-Rodriguez}}
\author[
   addressref={aff4},                   
]{\inits{K}\fnm{Klaus} \snm{Mauch}}
\author[
   addressref={aff1},                   
]{\inits{E}\fnm{Eva} \snm{Balsa-Canto}}
\author[
   addressref={aff3},                   
]{\inits{P}\fnm{Pedro} \snm{Mendes}}
\author[
   addressref={aff5,aff6},               
   email={yogi.jaeger@crg.eu}            
]{\inits{J}\fnm{Johannes} \snm{Jaeger}}
\author[
   addressref={aff1},                   
   corref={aff1},                       
   email={julio@iim.csic.es}            
]{\inits{JR}\fnm{Julio R} \snm{Banga}}

\address[id=aff1]{
  \orgname{Bioprocess Engineering Group, IIM-CSIC}, 
  \street{Eduardo Cabello 6},                       %
  \postcode{36208}                                  
  \city{Vigo},                                      
  \cny{Spain}                                       
}
\address[id=aff2]{%
  \orgname{European Molecular Biology Laboratory, European Bioinformatics Institute (EMBL-EBI)},
  \street{Wellcome Trust Genome Campus},
  \postcode{CB10 1SD}
  \city{Hinxton, Cambridge},
  \cny{UK}
}
\address[id=aff3]{%
  \orgname{School of Computer Science, Manchester Centre for Integrative Systems Biology, The University of Manchester},
  \street{131 Princess Street},
  \postcode{M1 7DN}
  \city{Manchester},
  \cny{UK}
}
\address[id=aff4]{%
  \orgname{Insilico Biotechnology AG},
  \street{Meitnerstra{\ss}e 8},
  \postcode{70563}
  \city{Stuttgart},
  \cny{Germany}
}
\address[id=aff5]{%
  \orgname{EMBL/CRG Research Unit in Systems Biology, Centre for Genomic Regulation (CRG)},
  \street{Dr. Aiguader 88},
  \postcode{08003}
  \city{Barcelona},
  \cny{Spain}
}
\address[id=aff6]{%
  \orgname{Universitat Pompeu Fabra (UPF)},
  \street{Pla\c{c}a de la Merc\'e, 10},
  \postcode{08002}
  \city{Barcelona},
  \cny{Spain}
}

\vspace{1cm}

\textbf{Abstract.} Dynamic modelling is one of the cornerstones of systems biology. Many research efforts are currently being invested in the development and exploitation of large-scale kinetic models. The associated problems of parameter estimation (model calibration) and optimal experimental design are particularly challenging. The community has already developed many methods and software packages which aim to facilitate these tasks. However, there is a lack of suitable benchmark problems which allow a fair and systematic evaluation and comparison of these contributions. Here we present BioPreDyn-bench, a set of challenging parameter estimation problems which aspire to serve as reference test cases in this area. This set comprises six problems including medium and large-scale kinetic models of the bacterium \textit{E. coli}, baker's yeast \textit{S. cerevisiae}, the vinegar fly \textit{D.~melanogaster}, Chinese Hamster Ovary cells, and a generic signal transduction network. The level of description includes metabolism, transcription, signal transduction, and development. For each problem we provide 
(i)		a basic description and formulation,
(ii)	implementations ready-to-run in several formats, 
(iii)	computational results obtained with specific solvers,
(iv)	a basic analysis and interpretation.
This suite of benchmark problems can be readily used to evaluate and compare parameter estimation methods. Further, it can also be used to build test problems for sensitivity and identifiability analysis, model reduction and optimal experimental design methods. The suite, including codes and documentation, can be freely downloaded from
\url{http://www.iim.csic.es/~gingproc/biopredynbench/}


\end{fmbox}
\end{frontmatter}

\section*{Background} 

Systems biology aims at understanding the organization of complex biological systems with a combination of mathematical modelling, experiments, and advanced computational tools. 
To describe the behaviour of complex systems, models with sufficient level of detail to provide mechanistic explanations are needed. 
This leads to the use of large-scale dynamic models of cellular processes \cite{link2014advancing}. 
By incorporating kinetic information, the range of applications of biological models can be widened.
The importance of kinetic models is being increasingly acknowledged in fields such as bioprocess optimization \cite{almquist2014kinetic}, metabolic engineering \cite{song2013modeling}, physiology, as well as cell and developmental biology \cite{jaeger2014bioattractors}.

Systems identification, or reverse engineering, plays an important part in the model building process. 
The difficult nature of reverse engineering was stressed in \cite{villaverde2014reverse}, where the different perspectives that coexist in the area of systems biology were reviewed.
Specifically, large-scale dynamic biological models generally have many unknown, non-measurable parameters. 
For the models to encapsulate as accurately as possible our understanding of the system (i.e. reproducing the available data and, ideally, being capable of making predictions), these parameters have to be estimated.
This task, known as parameter estimation, model calibration, or data fitting \cite{van2006dynamic,jaqaman2006linking,banga2008parameter,ashyraliyev2008systems,vanlier2013parameter}, consists of finding the parameter values that give the best fit between the model output and a set of experimental data. 
This is carried out by optimizing a cost function that measures the goodness of this fit. 
In systems biology models this problem is often multimodal (nonconvex), due to the nonlinear and constrained nature of the system dynamics. 
Hence, standard local methods usually fail to obtain the global optimum. 
As an alternative, one may choose a multistart strategy, where a local method is used repeatedly, starting from a number of different initial guesses for the parameters. 
However, this approach is usually not efficient for realistic applications, and global optimization techniques need to be used instead \cite{moles2003parameter,banga2008}.

Many methods have been presented for this task, but less effort has been devoted to their critical evaluation. It is clear, however, that to make progress in this research area it is essential to assess performance of the different algorithms quantitatively, in order to understand their weaknesses and strengths. Furthermore, if a new algorithm is to be accepted as a valuable addition to the state of the art, it must be first rigorously compared with the existing plethora of methods. 
This systematic comparison requires adequate benchmark problems, that is, reference calibration case studies of realistic size and nature that can be easily used by the community. Several collections of benchmarks -- and of methods for generating them -- have already been published \cite{mendes2003artificial,kremling2004benchmark,camacho2007comparison,gennemark2009benchmarks,haynes2009benchmarking,marbach2009generating,schaffter2011genenetweaver}. An artificial gene network generator, which allows to choose from different topologies, was presented in \cite{mendes2003artificial}. The system, known as A-BIOCHEM, generates pseudo-experimental noisy data in silico, simulating microarray experiments. An artificial gene network with ten genes generated in this way was later used to compare four reverse-engineering methods \cite{camacho2007comparison}. More recently, a toolkit called GRENDEL was presented with the same purpose \cite{haynes2009benchmarking}, including several refinements in order to increase the biological realism of the benchmark. A reverse-engineering benchmark of a small biochemical network was presented in \cite{kremling2004benchmark}. The model describes organism growth in a bioreactor and the focus was placed on model discrimination using measurements of some intracellular components. A proposal for minimum requirements of problem specifications, along with a collection of 44 small benchmarks for ODE model identification of cellular systems, was presented in \cite{gennemark2009benchmarks}. The collection includes parameter estimation problems as well as combined parameter and structure inference problems. Another method for generation of dynamical models of gene regulatory networks to be used as benchmarks is GeneNetWeaver \cite{schaffter2011genenetweaver}, which was used to provide the international Dialogue for Reverse Engineering Assessments and Methods (DREAM) competition with three network inference challenges (DREAM3, DREAM4 and DREAM5) \cite{marbach2009generating}. Subsequent competitions (DREAM6, DREAM7) included also parameter estimation challenges of medium-scale models \cite{meyer2014network}.
Similar efforts have been carried out in related areas, such as in optimization, where BBOB workshops (Black-Box Optimization Benchmarking, \cite{auger2012benchmarking}) have been organised since 2009. In this context it is also worth mentioning the collection of large-scale, nonlinearly constrained optimization problems from the physical sciences and engineering (COPS) \cite{dolan2004benchmarking}. 

Despite these contributions, there is still a lack of suitable benchmark problems in systems biology that are at the same time (i) dynamic, (ii) large-scale, (iii) ready-to-run, and (iv) available in several common formats. None of the above mentioned collections possesses all these features, although each one has a subset of them.
Here we present a collection of medium and large-scale dynamic systems, with sizes of tens to hundreds of variables and hundreds to thousands of estimated parameters, which can be used as benchmarks for reverse-engineering techniques. The collection includes two \textit{Escherichia coli} models \cite{chassagnole2002dynamic,kotte2010bacterial}, a genome-wide kinetic model of \textit{Saccharomyces cerevisiae} \cite{smallbone2013large}, a metabolic model of Chinese Hamster Ovary (CHO) cells \cite{villaverde2013highconfidence}, a signal transduction model of human cells \cite{macnamara2012state}, and a developmental gene regulatory network of \textit{Drosophila melanogaster} \cite{jaeger2004dynamic,crombach2012efficient,ashyraliyev2009gene}. 

Ensuring standardisation allows systems biology models to be reused outside of their original context: in different simulators, under different conditions, or as parts of more complex models \cite{krause2011sustainable}. Minimum requirements for published systems biology models are set out by the MIRIAM initiative \cite{novere2005minimum}: completeness of documentation, availability in standard formats, and semantic annotations connecting the model to web resources \cite{courtot2011controlled}. To this end, we have made five of the six models (the exception is the spatial model of \textit{D. melanogaster}) available in Systems Biology Markup Language (SBML \cite{hucka2003systems}) format, allowing for their simulation in multiple software tools, including AMIGO \cite{Balsa11} and COPASI \cite{hoops2006copasi}. Even when defined in a standard format such as SBML, large models such as the genome-wide kinetic model of \textit{S. cerevisiae} may give different results when simulated in different software environments. The inherent size and stiffness of genome-scale systems biology models create new challenges to be addressed for their robust simulation by systems biology tools \cite{smallbone2013large}.
To address this problem all the models have been consistently formatted, with their dynamics provided both in C and in Matlab. Additionally, a benchmark consisting of a parameter estimation problem has been defined for every model, for which ready-to-run implementations are provided in Matlab (optionally with the use of the AMIGO toolbox \cite{Balsa11}) and, in some cases, also in COPASI \cite{hoops2006copasi}. 
The availability of ready-to-run implementations is a highly desirable practice in computer science, since it ensures reproducibility of the results.
Calibration results with state of the art optimization methods are reported, which can serve as a reference for comparison with new methodologies. Additionally, suggestions on how to compare the performance of several methods are also given in the Results section.

\section*{Problem statement}

Given a model of a nonlinear dynamic system and a set of experimental data, the parameter estimation problem consists of finding the optimal vector of decision variables $p$ (unknown model parameters). This vector consists of the set of parameter values that minimize a cost function that measures the goodness of the fit of the model predictions with respect to the data, subject to a number of constraints. The output state variables that are measured experimentally are called observables. The following elements need to be clearly stated in order to properly define the calibration problem:

\begin{itemize}
\item  cost function to optimize (i.e. metric which reflects the mismatch between experimental and predicted values)
\item  dynamics of the systems (in our benchmark models they are given by systems of ordinary differential equations)
\item  model parameters to be estimated
\item  initial conditions for the dynamics (possibly unknown, in which case they are included among the parameters to be estimated)
\item  upper and lower bounds for the parameters
\item  state variables that can be measured (observed)
\item  values of external stimuli, also known as control variables
\item  measurements (over time and/or space) available for the calibration: number of experiments, stimuli (if any) for each experiment, data points per experiment, etc.
\item  (optional) type and magnitude of errors considered for the experimental data
\item  (optional) additional time series for cross-validation of the calibrated model
\item  solver used to numerically simulate the systems, and the relative and absolute error tolerances used
\end{itemize}

Mathematically, it is formulated as a nonlinear programming problem (NLP) with differential-algebraic constraints (DAEs), where the goal is to find $p$ to minimize an objective function. The objective function, or cost function, is a scalar measure of the distance between data and model predictions. There are several common choices for the objective function. The generalized least squares cost function is given by:
\begin{equation}\label{eq:statement}
J_{lsq} = \sum^{n_{\epsilon}}_{\epsilon=1} \sum^{n^{\epsilon}_o}_{o=1} \sum^{n^{\epsilon,o}_s}_{s=1} w^{\epsilon,o}_s \left(ym^{\epsilon,o}_s - y^{\epsilon,o}_s(p) \right)^2 
\end{equation}
where $n_{\epsilon}$ is the number of experiments, $n^{\epsilon}_o$ is the number of observables per experiment, and $n^{\epsilon,o}_s$ is the number of samples per observable per experiment. The measured data will be denoted as $ym^{\epsilon,o}_s$ and the corresponding model predictions will be denoted as $y^{\epsilon,o}_s(p)$. 
Finally, $w^{\epsilon,o}_s$ are scaling factors used to balance the contributions of the observables, according to their magnitudes and/or the confidence in the measurements. 
When information about the experimental error is available, one may use the maximum (log-)likelihood function to look for the parameters with the highest probability of generating the measured data. Assuming independently identically distributed measurements with normally distributed noise, the likelihood is defined as:

\begin{equation}\label{eq:llk}
J_{llk} = \sum^{n_{\epsilon}}_{\epsilon=1} \sum^{n^{\epsilon}_o}_{o=1} \sum^{n^{\epsilon,o}_s}_{s=1} \frac{ \left(ym^{\epsilon,o}_s - y^{\epsilon,o}_s(p) \right)^2 } {(\sigma^{\epsilon,o}_s)^2 }
\end{equation}

For known constant variances the log-likelihood cost function is similar to the generalized least squares, with weights chosen as the inverse of the variance, $w^{\epsilon,o}_s = 1/(\sigma^{\epsilon,o}_s)^2$. This is the case of most of the benchmark problems presented here (B1, B2, B4, B5). The exceptions are: problem B3, in which no noise has been added to the data, and thus the scaling factors $w^{\epsilon,o}_s$ are taken as the squared inverse of the maximum experimental value for each observable; and problem B6, where the weights are inversely related to the level of expression. More details are given in the next section. 
Note that the cost functions used in Matlab/AMIGO are not exactly the same as the ones used in COPASI, since in COPASI the weights are scaled so that for each experiment the maximal occurring weight is 1.

The minimization of the objective function is subject to the following constraints:
\begin{equation}\label{eq:constraints1}
\dot{x} = f\left(x,p,t \right)
\end{equation}
\begin{equation}\label{eq:constraints2}
x(t_0) = x_0
\end{equation}
\begin{equation}\label{eq:constraints3}
y = g(x,p, t)
\end{equation}
\begin{equation}\label{eq:constraints4}
h_{eq}(x,y,p) = 0
\end{equation}
\begin{equation}\label{eq:constraints5}
h_{in}(x,y,p) \leq 0
\end{equation}
\begin{equation}\label{eq:constraints6}
p^L \leq p \leq p^U
\end{equation}
where $g$ is the observation function, $x$ is the vector of state variables with initial conditions $x_0$, $f$ is the set of differential and algebraic equality constraints describing the system dynamics (that is, the nonlinear process model), $h_{eq}$ and $h_{in}$ are equality and inequality constraints that express additional requirements for the system performance, and $p^L$ and $p^U$ are lower and upper bounds for the parameter vector $p$.
The problem defined above is the general formulation of a nonlinear least squares optimization subject to dynamic constraints and bounds in the parameters. The problems included in this collection of benchmarks do not make use of constraints (\ref{eq:constraints4}--\ref{eq:constraints5}).

\subsection*{Remarks on parameter estimation methods}

Fitting a large, nonlinear model to experimental (noisy) data is generally a multimodal problem. In these circumstances, the use of local optimization methods, which are usually gradient-based, entails the risk of converging to local minima. Hence it is needed to use global optimization methods that provide more guarantees of converging to the globally optimal solution \cite{moles2003parameter,balsa2010}. Global optimization strategies can be roughly classified as deterministic, stochastic and hybrid. Deterministic methods can guarantee the location of the global optimum solution; however, their computational cost makes them unfeasible for large-scale problems. Stochastic methods, which are based on probabilistic algorithms, do not provide those guarantees, but are frequently capable of finding optimal or near-optimal solutions in affordable computation times. 

Some of the most efficient stochastic global optimization methods are the metaheuristic approaches.
A heuristic is an algorithm originated not from formal analysis, but from an expert knowledge of the task to be solved. A metaheuristic can be seen as a general-purpose heuristic method designed to guide an underlying problem-specific heuristic. It is therefore a method that can be applied to different optimization problems with few modifications. 
Hybrid methods which combine metaheuristics for global optimization and local methods for accelerating convergence in the vicinity of local minima can be particularly efficient. One such method is the enhanced Scatter Search algorithm, eSS \cite{Egea10}, and its parallel cooperative version, CeSS \cite{villaverde2012cooperative}. Matlab and R implementations are publicly available as part of the MEIGO toolbox \cite{Egea13}. The eSS method is available as a Matlab toolbox and is also included in AMIGO; this latter version is the one used in this work. It should be noted that AMIGO offers more than a dozen optimization solvers, including local and global methods, and the possibility of combining them to form user-defined sequential hybrid methods.
In COPASI \cite{hoops2006copasi} it is possible to choose among thirteen different optimization methods for parameter estimation, including deterministic and stochastic: Evolutionary Programming, Evolutionary Strategy (SRES), Genetic Algorithm, Hooke and Jeeves, Levenberg--Marquardt, Nelder--Mead, Particle Swarm, Praxis, Random Search, Simulated Annealing, Scatter Search, Steepest Descent, and Truncated Newton. 

\subsection*{Remarks on comparing optimization methods}

Although the objective of this paper is to present a set of ready-to-run benchmarks, we list below several guidelines on how to compare different optimizers with these problems.

Many optimization methods require an initial point and/or bounds on the decision variables. For ensuring a fair comparison between different methods, the same bounds and initial points should be set. Obviously, the nominal solution can not be used as an initial point. Special emphasis should be laid on ensuring full reproducibility. This entails providing all source codes and binary files used in computations, as well as specifying all implementation details, such as software and hardware environment (including compiler versions and options, if any). If some aspects of a method can be tuned, these settings must be clearly indicated.

Many different criteria may be used for comparing the performance of optimization methods. It can be expressed as a function of CPU time, number of function evaluations, or iteration counts. When considering several problems, a solver's average or cumulative performance metric can be chosen. If an algorithm fails to converge a penalty can be used, in which case an additional decision is required to fix its value. An alternative is to use ranks instead of numerical values, although this option hides the magnitudes of the performance metric. All approaches have advantages and drawbacks, and their use requires making choices that are subjective to some extent. In an attempt to combine the best features of different criteria, Dolan and Mor\'e \cite{dolan2002benchmarking} proposed to compare algorithms based on their performance profiles, which are cumulative distribution functions for a performance metric. Performance profiles basically rely on calculations of the ratio of the solver resource time versus the best time of all the solvers. It should be noted that, for complex large-scale problems where identifiability is an issue, different methods often arrive at different solutions. In that case the use of performance profiles requires choosing a tolerance to define acceptable solutions. Performance profiles are a convenient way of summarizing results when there are many methods to be compared and many problems on which to test them. When this is not the case, however, more information can be provided by using convergence curves. Convergence curves plot the evolution of the objective function, as defined in Equation (\ref{eq:statement}), as a function of the number of evaluations or  the computation time (since the overhead is different for each method). They provide information not only about the final value reached by an algorithm, but also about the speed of progression towards that value. 

When comparing different optimization methods, the best result (cost) and the mean (or median) for N runs should be reported in a table. Similar statistics for computation and number of evaluations should apply to all the methods. However, since the final values can be greatly misleading, convergence curves should be provided in addition to this table.

Note that, in order to make a fair comparison of convergence curves obtained with different software tools and/or hardware environments, it is a good practice to report any speedup due to parallelism. This can happen in non-obvious situations. For example, COPASI can make use of several threads in multi-core PCs due to its use of the Intel MKL library. In summary, fair comparisons should be made taking into account the real overall computational effort used by each method/implementation. 

\subsection*{Remarks on identifiability}

Parameter estimation is just one aspect of what is known as the inverse problem. This larger problem also includes identifiability analysis, which determines whether the unknown parameter values can be uniquely estimated \cite{walter1997identification,gadkar2005iterative}. Lack of identifiability means that there are several possible parameter vectors that give the same agreement between experimental data and model predictions. 
We may distinguish between a priori structural identifiability and a posteriori or practical identifiability \cite{walter1997identification,gadkar2005iterative}. The parameters are structurally identifiable if they can be uniquely estimated from the designed experiment under ideal conditions of noise-free observations and error-free model structure. Structural identifiability is a theoretical property of the model structure, which can be very difficult to determine for large and complex models \cite{Chis11a, Chis11b, becker2013reverse}. Even if a model is structurally identifiable, it may exhibit practical identifiability issues. Practical identifiability depends on the output sensitivity functions (partial derivatives of the measured states with respect to the parameters). If the sensitivity functions are linearly dependent the model is not identifiable, and sensitivity functions that are nearly linearly dependent result in parameter estimates that are highly correlated. 
Furthermore, even if they are linearly independent, low sensitivities may lead to an undesirable situation.
Practical identifiability can be studied from sensitivity-based criteria like the Fisher information matrix (FIM). The practical identifiability of the models can be analyzed in this way with the AMIGO toolbox \cite{Balsa11}. The AMIGO\_LRank method ranks the model parameters according to their influence on the model outputs, using several sensitivity measures.
In large biological models identifiability issues are the norm rather than the exception \cite{zak2003importance,yue2006insights,anguelova2007conservation,ashyraliyev2008systems,srinath2010parameter,Gabor2011,miao2011identifiability,Chis11a,jia2012incremental,cedersund2012conclusions,berthoumieux2013identifiability,distefano2014dynamic}. This may be partly due to inconsistent modelling practices, but even when a model has been carefully built and is structurally identifiable, the amount of data required for a perfect calibration (practical identifiability) is usually large. As an illustration, consider the general case of a model described by differential equations, and assume the ideal situation where the structure of the equations is perfectly known. Then, a well-known result states that identification of $r$ parameter values requires $2r+1$ measurements \cite{Sontag02}. However, it is frequently the case that a model with more than a thousand parameters has to be calibrated with only dozens or maybe hundreds of measurements.
Finally, it should be noted that lack of identifiability does not preclude the use of model-based methods. Unique model predictions can in fact be obtained despite unidentifiability, as discussed by Cedersund \cite{cedersund2012conclusions}.

\section*{Benchmark problems}

Here we present a collection of parameter estimation problems and their descriptions. The characteristics of the six dynamic models are summarized in Table \ref{tab:models}. Four of the benchmark problems have been defined using in silico experiments, where pseudoexperimental data have been generated from simulations of the models and addition of artificial noise. The use of simulated data is usually considered the best way of assessing performance of parameter estimation methods, because the true solution is known. Additionally, we provide two benchmark problems that use real data. For each problem we provide the following information (see Supplementary Information online, \small{\url{http://www.iim.csic.es/~gingproc/biopredynbench/}}):

\begin{itemize}
\item Dynamic model.
\item Experimental information: initial conditions, input functions, what is measured, measurement times, noise level and type.
\item Cost function to be used: its type (least squares, weighted least squares, maximum likelihood, etc), and why it should be chosen.
\item Parameters to estimate: lower and upper bounds, initial guesses, nominal values (the latter must not be used during estimations).
\item Implementations (Matlab with and without the AMIGO toolbox, C, COPASI): installation, requirements, and usage. Ready-to-run scripts are provided, with examples of how to execute them and their expected output.
\end{itemize}

\begin{table*}[t]
\begin{flushleft}
  \caption{
  \textbf{Models}
  }
  \begin{tabular}{|l|l|l|l|l|l|l|}
  \hline
Model ID           & \textbf{B1}              & \textbf{B2}                  & \textbf{B3}             & \textbf{B4}  & \textbf{B5}  & \textbf{B6} \\
\hline
Model Ref          &\cite{smallbone2013large} &\cite{chassagnole2002dynamic} &\cite{kotte2010bacterial}&\cite{villaverde2013highconfidence}&\cite{macnamara2012state}&\cite{crombach2012efficient}\\
Cell               & \textit{S. cerevisiae}   & \textit{E. coli}             & \textit{E. coli}        & CHO          & Generic      & \textit{Drosophila}\\        
                   &                          &                              &                         &              &              & \textit{melanogaster}\\ 
Description        & Metabolic:               & Metabolic:                   & Metabolic: CCM          & Metabolic    & Signal       & Developmental \\
level              & genome scale             & CCM                          & \& transcription        &              & transduction & GRN (spatial) \\
Parameters         & 1759                     & 116                          & 178                     & 117          & 86           & 37 \\
Dynamic states     & 276                      & 18                           & 47                      & 34           & 26           & 108 --212 \\
Observed states    & 44                       &  9                           & 47                      & 13           & 6            & 108 --212 \\
Experiments        &  1                       & 1                            &  1                      &   1          & 10           & 1 \\
Data points        & 5280                     & 110                          & 7614                    & 169          & 96           & 1804 \\      
Data type          & simulated                & measured                     & simulated               & simulated    & simulated    & measured \\
Noise level        & $20\%$                   & real                         & no noise                & variable     & $5\%$        & real \\
\hline
\end{tabular}
\\
Main features of the benchmark models.
\end{flushleft}
\label{tab:models}
\end{table*}

\subsection*{Problem B1: genome-wide kinetic model of  \textit{S. cerevisiae}}

The biochemical structure of this model is taken from yeast.sf.net (version 6, \cite{Heavner12}). In decompartmentalised form, this network has 1156 reactions and 762 variables. We fix some experimentally determined exchange fluxes, and use geometric FBA \cite{Smallbone09} to choose a unique reference flux distribution consistent with the experimental data. 
We fix some initial concentrations to their experimentally determined levels and assign the remainder typical values. We define reaction kinetics using the common modular rate law, a generalised form of the reversible Michaelis-Menten kinetics that can be applied to any reaction stoichiometry \cite{liebermeister2010modular}. 
The final model contains 261 reactions with 262 variables and 1759 parameters. 
This model has been created according to the pipeline presented in \cite{smallbone2013large}, which ensures consistency with our sparse data set; whilst no data is required to produce the model, it can incorporate any known flux or concentration data or any kinetic constants. As an addition to the model developed in \cite{smallbone2013large}, this version has been alligned with previously unpublished experimental data. The new data consist of 44 steady-state measurements (38 concentrations and 6 fluxes), which are included in Tables 14 and 15 of the Supplementary Information online. The steady state is found to be stable. The number of measurements available at the present stage is not enough for carrying out a proper model calibration. Envisioning that dynamic (time-series) measurements of the 44 observed variables may be available in the near future, we show in this paper how they will be employed for re-estimating the parameter values. With this aim, we have generated pseudo-experimental noisy data corresponding to a pulse in the concentration of extracellular glucose, and have used this simulated data to re-calibrate the model. We generated 120 samples per observable and added artificial measurement noise (20\%) to resemble realistic conditions.

\subsection*{Problem B2: dynamic model of the Central Carbon Metabolism of \textit{E. coli}}

This model, originally published in \cite{chassagnole2002dynamic} and available at the BioModels database \cite{BioModels2010}, reproduces the response to a pulse in extracellular glucose concentration. It includes 18 metabolites in two different compartments: the cytosol (17 internal metabolites), and the extracellular compartment (1 extracellular metabolite: glucose). These metabolites are involved in 48 reactions: 30 kinetic rate reactions, 9 degradation equations, 8 dilution equations, and 1 equation for extracellular glucose kinetics. Additionally, there are 7 analytical functions, thus the model is defined by a total of 55 mathematical expressions. We have reformulated the model to use it as a parameter estimation problem; the 116 parameters to be estimated consist of kinetic parameters and maximum reaction rates.
As an addition to the model version available in the Biomodels database, we provide the experimental data that were used in the original publication but had not been published (Klaus Mauch, personal communication). The dataset is given in Table 16 of the Supplementary Information online, and consists of time-course concentration measurements of nine metabolites. The aim of the model calibration in this case is to find a better fit to the experimental data than the one obtained with the nominal parameter vector used in the original publication \cite{chassagnole2002dynamic}. Note that this is different from benchmarks 1 and 3--5, which use simulated data and where the aim is to recover a fit as good as the one obtained with the nominal parameter vector, with which the data were generated.

\subsection*{Problem B3: enzymatic and transcriptional regulation of the Central Carbon Metabolism of \textit{E. coli}}

This model simulates the adaptation of \textit{E. coli} to changing carbon sources. Complete information about this model is available as the supplementary information of \cite{kotte2010bacterial}. It is also included in the BioModels Database \cite{BioModels2010}. It should be noted that there are some differences in parameter values between the original model and the BioModels version; however, these changes do not alter the simulation results, a fact that indicates unidentifiability. The model contains 47 ODEs and 193 parameters, of which 178 are considered unknown and need to be estimated. The other 15 parameters are constants known to the modeler (number of subunits of the multimers--enzymes--, scaling factors, universal protein degradation rate, and gene expression rate constant). 
The outputs of the system are the 47 state variables, which represent concentrations. Pseudo-experimental data were generated by simulation of the sixth scenario defined in the simulation files included as supplementary material in \cite{kotte2010bacterial}. This scenario simulates an extended diauxic shift which consists of three consecutive environments, where the carbon sources are first glucose, then acetate, and finally a mixture of both. Under these conditions, the 47 concentration profiles are sampled every 1000 seconds, for a total of 162 time points (45 hours). This model exhibits large differences in value among concentrations, which span five orders of magnitude. To equalize their contribution to the objective function, we scale each time-series dividing it by the maximum of the experimental value (scaled least squares).

\subsection*{Problem B4: metabolic model of Chinese Hamster Ovary (CHO) cells}

Chinese Hamster Ovary cells (CHO) are used for protein production in fermentation processes \cite{wurm2004production}. This model simulates a batch process with resting cells: no metabolites are fed for a final time horizon of 300 hours. The fermenter medium contains glucose as main carbon source, and leucine and methionine are the main amino acids taken up. Lactate was modelled to be a by-product of the fermentation process. A generated protein serves as main product of the fermentation process. The model comprises 35 metabolites in three compartments (fermenter, cytosol, and mitochondria) and 32 reactions, including protein product formation, Embden-Meyerhof-Parnas pathway (EMP), TCA cycle, a reduced amino acid metabolism, lactate production, and the electron transport chain. The kinetics are modelled as in \cite{chassagnole2002dynamic}, and the resulting ODE model comprises 117 parameters in total. Some aspects of this model were partially discussed in \cite{villaverde2013highconfidence}. For optimization purposes pseudo-experimental data were generated, mimicking a typical cell behavior. The following 13 metabolites are assumed to be measured: in fermenter, glucose, lactate, product protein, leucine, and methionine; in cytosol, aspartate, malate, pyruvate, oxaloacetate, ATP, and ADP; and in mitochondria, ATP and ADP. Samples were assumed to be daily taken over the whole fermentation time. 

\subsection*{Problem B5: signal transduction logic model}

To illustrate the advantages and disadvantages of different formalisms related to logic models, MacNamara and colleagues constructed a plausible network of interactions consisting of signaling factors known to be activated downstream of $EGF$ and $TNF\text{-}\alpha$ \cite{macnamara2012state}. The model consists of 26 ODEs that use a logic-based formalism, which is explained in detail in \cite{wittmann2009transforming}. In this formalism, state values can vary between 0 and 1 and represent the normalized activity of a given protein, which is typically measured as the level of phosphorylation. In total the model includes 86 continuous parameters, corresponding to the half maximal activations ($k)$, the Hill coefficients ($n$) and a set of parameters controlling the rate of activation/deactivation of a given protein ($\tau$). The model incorporates $EGF$ and $TNF\text{-}\alpha$ which are treated as stimuli that trigger the pathway response. In addition to these two stimuli, the model includes two kinase inhibitors for $RAF1$ and $PI3K$, which can block the activity of both species. In total the model can be perturbed by these 4 cues, allowing a rich variation in the dynamic profiles of the model signaling components, an essential requirement for parameter estimation. In order to generate a data-set for reverse engineering the model structure, the authors generated data, corresponding to 10 in-silico experiments, where the different cues (stimuli and inhibitors) are added in different combinations. For each experiment 6 strategically located states are observed. Each observable was measured at 16 equidistant time points per experiment. In addition to this Gaussian noise was added to the data in order to mimic a reasonable amount of experimental error.
Note that the SBML implementation of this model uses the SBML qual format \cite{chaouiya2013sbml}, an extension of SBML developed for qualitative models of biological networks.

\subsection*{Problem B6: the gap gene network of the vinegar fly, \textit{Drosophila melanogaster}}

Our last benchmark model is slightly different from those previously described, in that it represents a spatial model of pattern formation in multi-cellular animal development, and the data for fitting are based on microscopy, rather than metabolomics or transcriptomics.
The gap genes form part of the segmentation gene network, which patterns the anterior--posterior (AP) axis of the \textit{Drosophila melanogaster} embryo. They are the primary regulatory targets of maternal morphogen gradients, and are active during the blastoderm stage in early development. 
In the model, the embryo is a single row of dividing nuclei along the AP axis, with each nucleus containing the four gap genes and receiving input from four external factors. The gap genes included in the model are \textit{hunchback} (\textit{hb}), \textit{Kr\"uppel} (\textit{Kr}), \textit{giant} (\textit{gt}), and \textit{knirps} (\textit{kni}), and the external inputs Bicoid (Bcd), Caudal (Cad), Tailless (Tll), and Huckebein (Hkb). Three processes occur within and between nuclei: (1) regulated gene product synthesis, (2) Fickian gene product diffusion, and (3) linear gene product decay. These processes are formalised with ODEs, and result in the model having 37 unknown parameters. 
This model \cite{crombach2012efficient,ashyraliyev2009gene} implements the gene circuit approach \cite{Mjolsness1991connectionist,reinitz1995mechanism} used to reverse-engineer the regulatory interactions of the gap genes by fitting to quantitative spatio-temporal gene expression data, which can be mRNA \cite{crombach2012efficient} or protein \cite{ashyraliyev2009gene}. 
The data consist of 9 time points spanning 71 minutes of \textit{Drosophila} development, and at each time point maximally 53 nuclei with data points for the four gap genes, and the four external inputs.
The fit is measured with a weighted least squares scheme (WLS) with variable weights, which, in the case of the mRNA data used here \cite{crombach2012efficient}, are inversely related to the level of expression. The weights were created from normalized, integrated mRNA expression data according to the formula: $w = 1.0 - 0.9y$, with $y\in[0,1]$ being the normalized staining intensity. This proportionality of variation with expression level reflects the fact that gap domains (showing high levels of expression) show more variation than those regions of the embryo in which a gene is not expressed \cite{crombach2012efficient}.

\section*{Results}

We show how our collection of benchmark problems can be used by reporting selected results using several parameter estimation methods. We emphasize that the purpose of this work is not to provide a comprehensive comparison of all existing approaches, but to provide a useful, versatile, and practical test set and illustrate its use. For simplicity, and to enable direct comparisons among benchmarks, all the computations reported in this section have been carried out in Matlab, using the algorithms available in the AMIGO toolbox \cite{Balsa11}. This includes both global and local optimization methods; the latter have been used in a multistart procedure, where multiple instances are launched from different initial points selected randomly within the parameter bounds. 

\begin{figure*}[ht]
\begin{center}
\includegraphics[width=12cm]{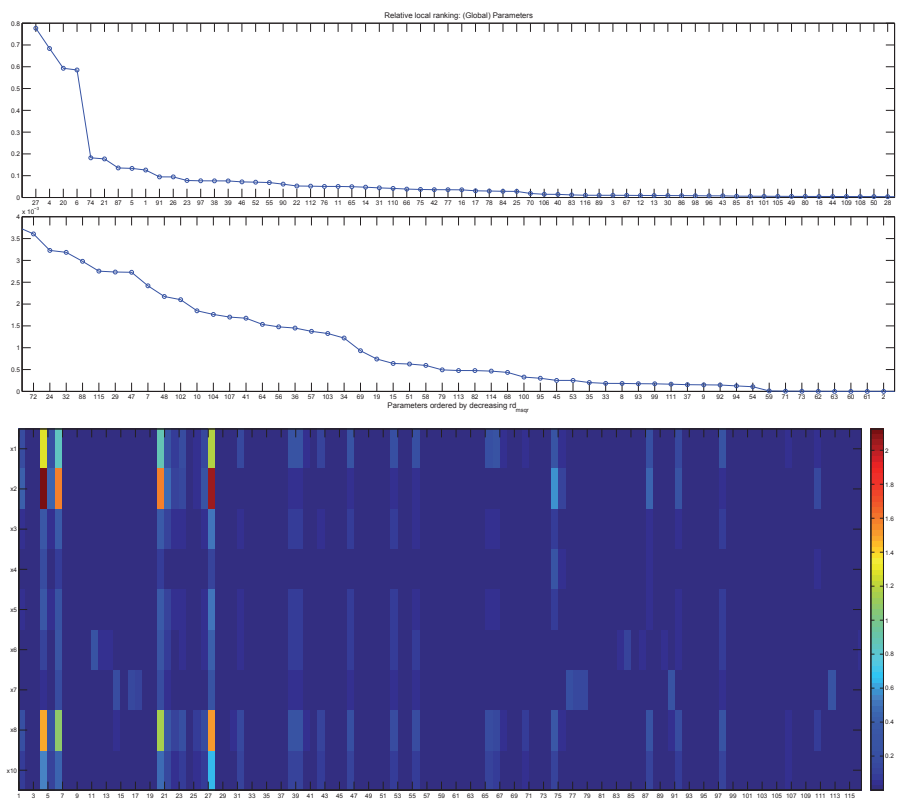}
\end{center}
\caption{\textbf{Benchmark 2: sensitivities.} The two panels on top show the local rank of the parameters, i.e., the parameters ordered in decreasing order of their influence on the system's behaviour ($\delta_p^{msqr}$). Note that the middle panel is a continuation of the upper one with a smaller y-axis scale.
The array in the bottom panel shows the sensitivity of the 9 state variables (metabolite concentrations, in columns) of the model with respect to the 116 parameters. The colour bar in the right shows the sensitivity range: high sensitivities are plotted in red, low sensitivities in blue.}
\label{b2_fs1}
\end{figure*}

Before estimating the parameter values we assessed the identifiability of the models. Model parameters were ranked according to their influence on the system output (sensitivity), using the local rank routine (AMIGO\_LRank) from the AMIGO toolbox as described in the previous section. As is typical of models of this size, it was found that all benchmarks have identifiability issues, with a portion of their parameters exerting very little influence on the model outputs. Therefore, the goal of these benchmarks is not to obtain accurate estimates of all the parameters, but rather to obtain a good fit to the data: when tested on this collection of benchmarks, optimization methods should be evaluated by their ability to minimize the objective function.
As an illustration of the typical outcome that can be obtained from the local rank method, we show in Figure \ref{b2_fs1} the results of the practical identifiability analysis for problem B2. Figure \ref{b2_fs1} ranks the parameters in decreasing order of their influence on the system's behaviour, which is quantified by means of the importance factors $\delta_p^{msqr}$:
\begin{equation}
\delta_p^{msqr} = \frac{1}{n_{lhs}n_d}\sqrt{\sum_{mc=1}^{n_{lhs}}\sum_{d=1}^{n_{d}}([s_d]_{mc})^2}
\end{equation}
\noindent where $n_{lhs}$ are the different values for each of the parameters selected by Latin Hypercube Sampling, $n_d$ is the number of experiments, and $[s]$ are the relative sensitivities.
Figure \ref{b2_fs1} shows the sensitivity of the state variables with respect to the parameters. From this figure it becomes clear that many parameters such as 8-10, 32-38, 56-64, are not influencing observables. Therefore those parameters are expected to be poorly identifiable.

In the remainder of this section we show selected results of the best performing optimization methods in every parameter estimation problem. Complete results for every benchmark are reported in the Supplementary Information online. 

\begin{figure}[ht]
\begin{center}
\includegraphics[width=6cm]{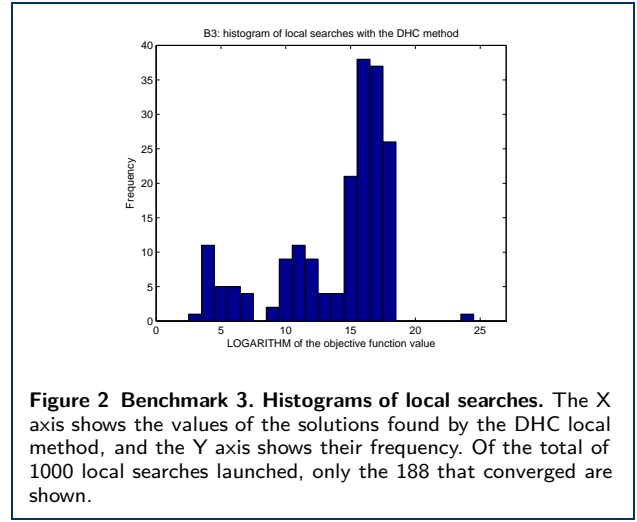} 
\end{center}
\caption{\textbf{Benchmark 3. Histograms of local searches.} The X axis shows the values of the solutions found by the DHC local method, and the Y axis shows their frequency. Of the total of 1000 local searches launched, only the 188 that converged are shown.}
\label{MULTISTARTS}
\end{figure}

To evaluate the performance of local methods we launched repeated local searches in a multistart procedure, starting from initial parameter vectors with values chosen randomly from within the parameter bounds. 
It should be noted that, while multistarts of local searches are a popular option for parameter estimation, they are usually not the most efficient solution when dealing with large-scale nonlinear models. Due to the multimodal nature of these problems, local methods tend to be stuck in local minima, which can sometimes be very far from the global optimum. Launching local methods from random points leads to spending a large fraction of the computational time in unsuccessful searches. Hence, global optimization methods usually perform better in these cases, especially if--as happens with eSS--they are used in combination with local searches.
As an example, Figure \ref{MULTISTARTS} shows histograms of the results (i.e., objective function values reached and the frequency with which they were found) obtained with the DHC local method for benchmark B3. Similar outcomes were obtained with the other benchmarks and methods. Complete results for all the benchmarks and with different methods are included in the Supplementary Information online.
In all cases, the number of local searches was fixed so that their overall CPU time was comparable to that consumed in optimizations where the global method eSS was used. While there was great variability in the results obtained for the different benchmarks, a conclusion was common to all of them: in all cases, the local methods were outperformed by the global optimization method eSS.

\begin{figure*}[ht]
\begin{center}
\includegraphics[width=12cm]{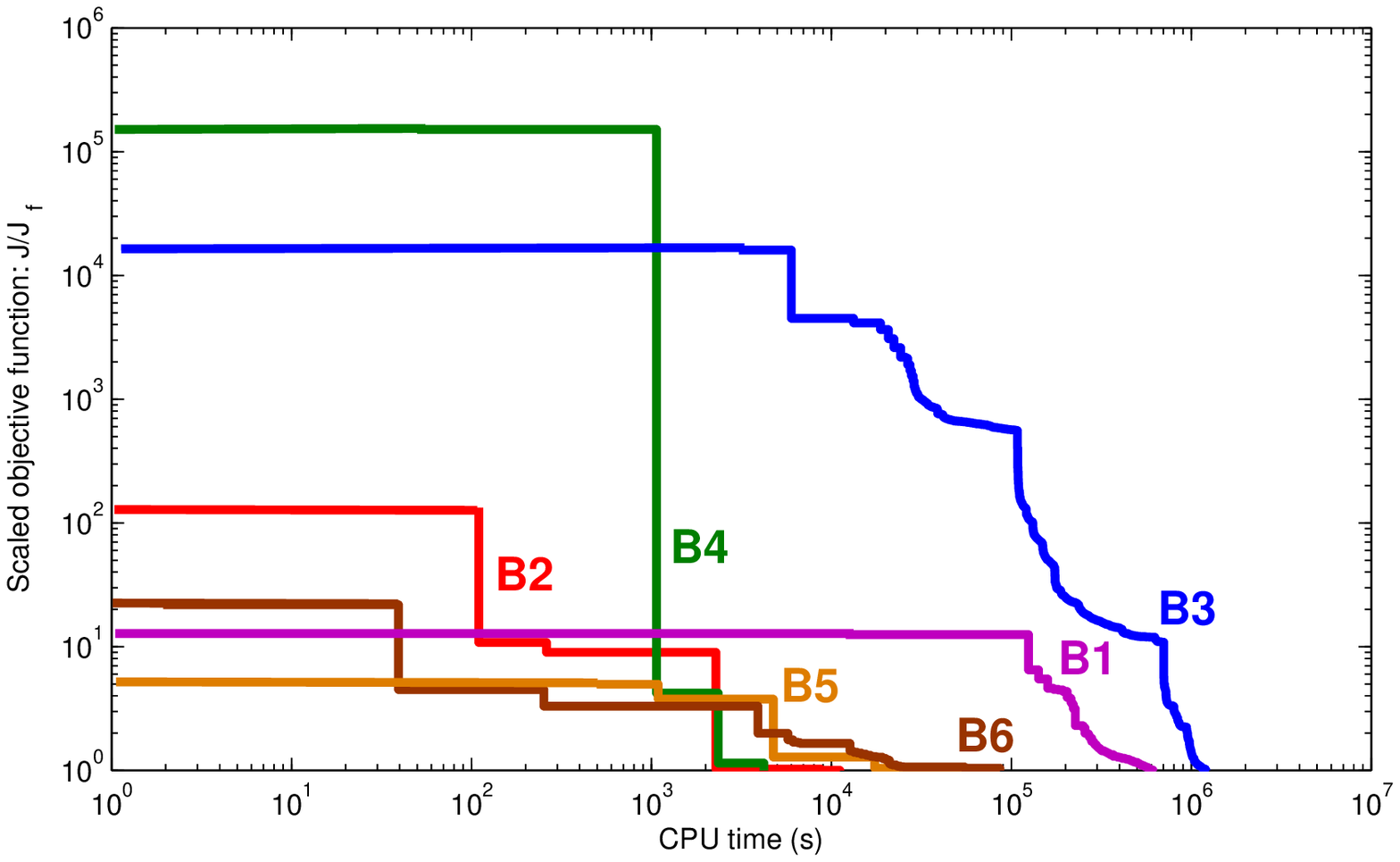}
\end{center}
\caption{\textbf{Convergence curves.} Representative results of parameter estimation runs of the six benchmarks, carried out with the eSS method. The curves plot the (logarithmic) objective function value as a function of the (logarithmic) computation time. 
For ease of visualization, the values in the curves have been divided by the final value reached by each of them, i.e. the $y$ axis plots $J/J_f$.
Note that, since the benchmarks have different number of variables and data points, and different noise levels, the objective function values are not equivalent for different models. 
Results obtained on a computer with Intel Xeon Quadcore processor, 2.50 GHz, using Matlab 7.9.0.529 (R2009b) 32-bit.}
\label{convergence}
\end{figure*}

\begin{figure*}[ht]
\begin{center}
\includegraphics[width=12cm]{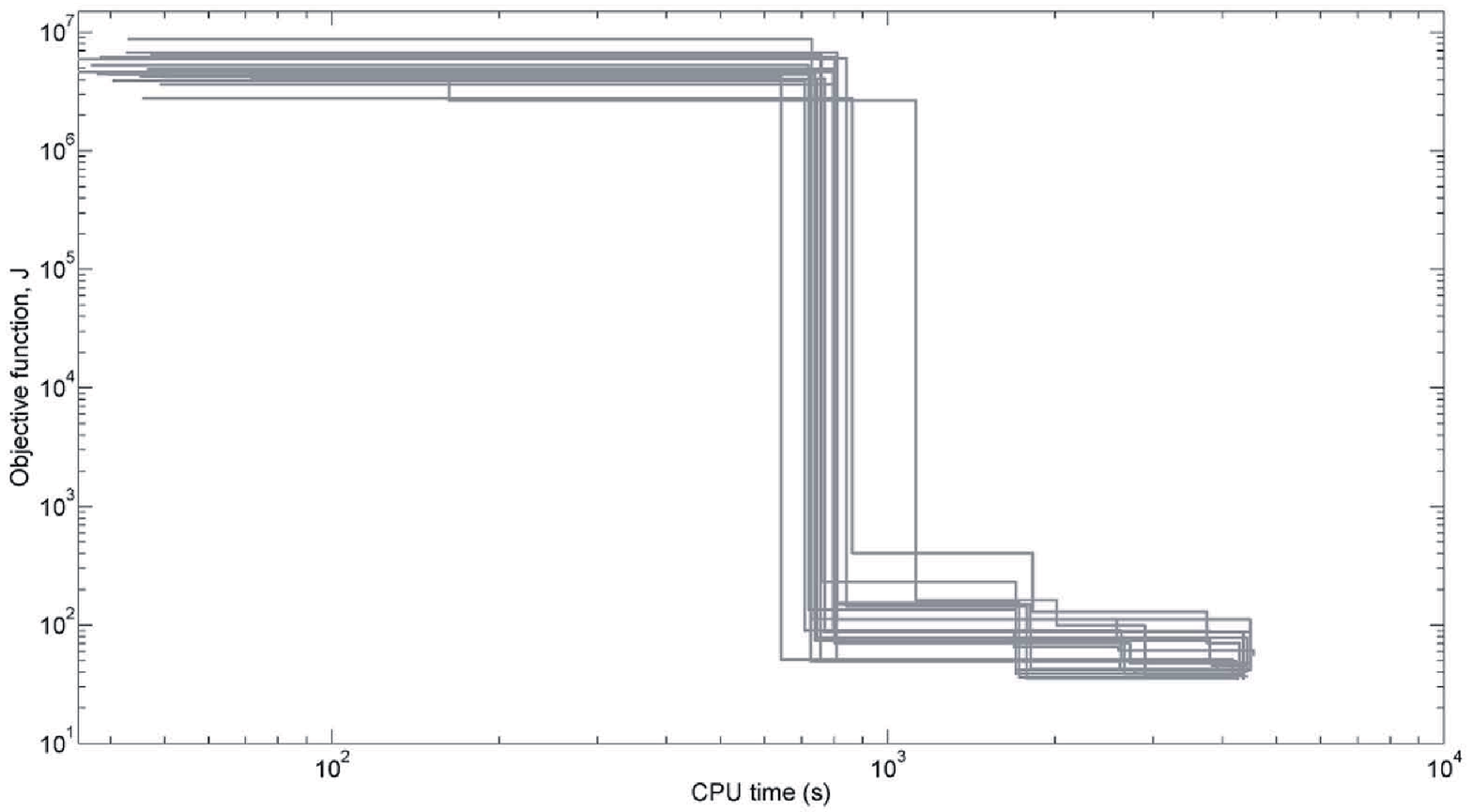}
\end{center}
\caption{\textbf{Dispersion of convergence curves.} Results of 20 parameter estimation runs of the B4 benchmark (CHO cells) with the eSS method. The figures plot the objective function value as a function of the computation time (in log-log scale). Results obtained on a computer with Intel Xeon Quadcore processor, 2.50 GHz, using Matlab 7.9.0.529 (R2009b) 32-bit.}
\label{conv_CHO}
\end{figure*}

The convergence curves of the six benchmarks are shown in Figure \ref{convergence}. Results were obtained with the eSS method on a computer with Intel Xeon Quadcore processor, 2.50 GHz. It can be clearly noticed that, due to the differences in size and complexity, the computational cost of estimating the parameters varies among benchmarks. Results show that they can be naturally classified in three different levels: 
\begin{itemize}
\item B1 and B3 are the most expensive: in our computers, obtaining a reasonably good fit took at least one week.
\item B5 and B6 are intermediate in terms of cost; a good fit could be obtained in one day.
\item B2 and B4 are the least expensive, with good fits obtained in one or a few hours.
\end{itemize}
These computation times can be used as a reference to select the appropriate benchmarks to test a particular optimization method, depending on its focus and the available time.
Due to the stochastic nature of the eSS algorithm, results may vary among optimization runs. Figure \ref{conv_CHO} shows the dispersion of 20 different optimization results for benchmark B4.

Table \ref{tab:pe} summarizes the settings and outcomes of the parameter estimations with eSS, including the local method used for each problem. Note that, while DN2FB is generally recommended \cite{rodriguez2006novel}, we have realized that for large-scale problems it may not be the most efficient local method, due to the large number of evaluations needed to calculate the derivatives. Hence, for the problems considered here it is outperformed by other methods like DHC, SOLNP, or FMINCON. 

One of the outcomes reported in Table \ref{tab:pe} is the cumulative normalized root-mean-square error, $\sum{}$NRMSE. The root-mean-square error is a standard measure of the goodness of fit obtained for an observable which is defined as
\begin{equation}
\mathrm{RMSE} = \sqrt{ \frac{  \sum^{n_{\epsilon}}_{\epsilon=1} \sum^{n^{\epsilon,o}_s}_{s=1} \left(ym^{\epsilon,o}_s - y^{\epsilon,o}_s(p) \right)^2 } {n_{\epsilon} \cdot n^{\epsilon,o}_s } }
\end{equation}
with the same notation as in equation (\ref{eq:statement}). To account for the different magnitudes of the observables it is useful to report the normalized root-mean-square error, NRMSE, which scales the NRMSE by dividing it by the range of values of the observable:
\begin{equation}\label{eq:NRMSE}
\mathrm{NRMSE} = \frac{\mathrm{RMSE}}{max( ym^{\epsilon,o}  )-min( ym^{\epsilon,o} )}
\end{equation}
The cumulative normalized root-mean-square error, $\sum{}$NRMSE, is simply the sum of the NRMSE for all observables.

\begin{table*}[t]
\begin{flushleft}
  \caption{
  \textbf{Parameter estimation with eSS (AMIGO implementation): settings and results}
  }
  \begin{tabular}{|l|l|l|l|l|l|l|}
  \hline	
Model ID            & \textbf{B1}          & \textbf{B2}             & \textbf{B3}              & \textbf{B4}             & \textbf{B5}           & \textbf{B6}  \\
\hline
$p^U$               & $5\cdot p_{nom}$     & $10\cdot p_{nom}^{(ex)}$  & $10\cdot p_{nom}^{(ex)}$  & $5\cdot p_{nom}$     & varying               & varying      \\
$p^L$               & $0.2\cdot p_{nom}$   & $0.1\cdot p_{nom}^{(ex)}$ & $0.1\cdot p_{nom}^{(ex)}$ & $0.2\cdot p_{nom}$   & varying               & varying      \\	
Local method        & DHC                  & FMINCON                 & none                     & FMINCON                 & DHC                   & FMINCON      \\	
CPU time            & $\approx$170 hours   & $\approx$3 hours        & $\approx$336 hours       & $\approx$1 hour         & $\approx$16 hours     & $\approx$24 hours  \\
Evaluations         & $6.9678\cdot10^5$    & $9.0728\cdot10^4$       & $7.2193\cdot10^6$        & $1.6193\cdot10^5$       & $8.8393\cdot10^4$     & $2.0751\cdot10^6$ \\
$J_0$               & $5.8819\cdot10^9$    & $3.1136\cdot10^4$       & $4.6930\cdot10^{16}$     & $6.6034\cdot10^8$       & $3.1485\cdot10^4$     & $8.5769\cdot10^5$ \\
$J_f$	              & $1.3753\cdot10^4$    & $2.3390\cdot10^2$       & $3.7029\cdot10^{-1}$     & $4.5718\cdot10^1$       & $3.0725\cdot10^3$     & $1.0833\cdot10^5$ \\
$J_{nom}$           & $1.0846\cdot10^6$    & $-$                     & $0$                      & $3.9068\cdot10^1$       & $4.2737\cdot10^3$     & $ -$ \\
$\sum$NRMSE$_0$     & $3.5834\cdot10$      & $8.5995\cdot10^{-2}$    & $3.5457\cdot10$          & $4.8005\cdot10$         & $4.0434\cdot10^1$     & $2.3808\cdot10^2$ \\	
$\sum$NRMSE$_f$     & $5.7558$             & $2.4921$                & $2.9298\cdot10^{-1}$     & $2.8010$                & $2.7430\cdot10^1$     & $1.6212\cdot10^2$ \\
$\sum$NRMSE$_{nom}$ & $3.8203$             & $-$                     & $0$                      & $2.8273$                & $3.0114\cdot10^1$     & $ -$ \\	
\hline
\end{tabular}
\\
Optimization settings and results obtained for each of the benchmarks with the eSS method, using the implementation provided in the AMIGO toolbox. In some cases the lower ($p^L$) and upper ($p^U$) bounds in the parameters are specified as a function of the nominal parameter vector, $p_{nom}$. There may be exceptions to these bounds, in cases where it makes sense biologically to have a different range of values (e.g. Hill coefficients in the range of 1--12). Cases with exceptions are marked by $^{(ex)}$. In other cases all the parameters have specific bounds; this is marked as ``varying''. The initial objective function value, $J_0$, corresponds to the parameter vector $p_0$ used as initial guess in the optimizations, which is randomly selected between the bounds $p^L$ and $p^U$. The only exception is benchmark B2, where $p_0$ is the parameter vector reported in the original publication. 
The final value achieved in the optimizations is $J_f$. $\sum$NRMSE is the cumulative normalized root-mean-square error as defined in eq. (\ref{eq:NRMSE}).
Results obtained on a computer with Intel Xeon Quadcore processor, 2.50 GHz, using Matlab 7.9.0.529 (R2009b) 32-bit.
\end{flushleft}
\label{tab:pe}
\end{table*}

Note that, due to the realistic nature of most of these problems, there may be lack of identifiability and optimization may result in overfitting: that is, an optimal solution may be found that gives a better fit to the pseudoexperimental data than the one obtained with the nominal parameter vector used to generate the data. This is explained because, in the presence of measurement noise, the optimal solution manages to fit partially not only the system dynamics, but also the noise itself--which of course cannot be achieved by the nominal solution. Hence in the results reported in Table \ref{tab:pe} the optimal objective function value ($J_f$) is sometimes smaller (i.e. better) than the nominal one ($J_{nom}$).
This may also happen with the NRMSE values. Note however that, since the objective functions used in the calibration ($J$) and the NRMSE are different metrics, their behavior may be different. For example, for B1 $J_f<J_{nom}$ and NRMSE$_f>$NRMSE$_{nom}$, while for B4 the opposite is true: $J_f>J_{nom}$ and NRMSE$_f<$NRMSE$_{nom}$.

\begin{center}
\begin{figure*}[ht]
\includegraphics[width=14cm]{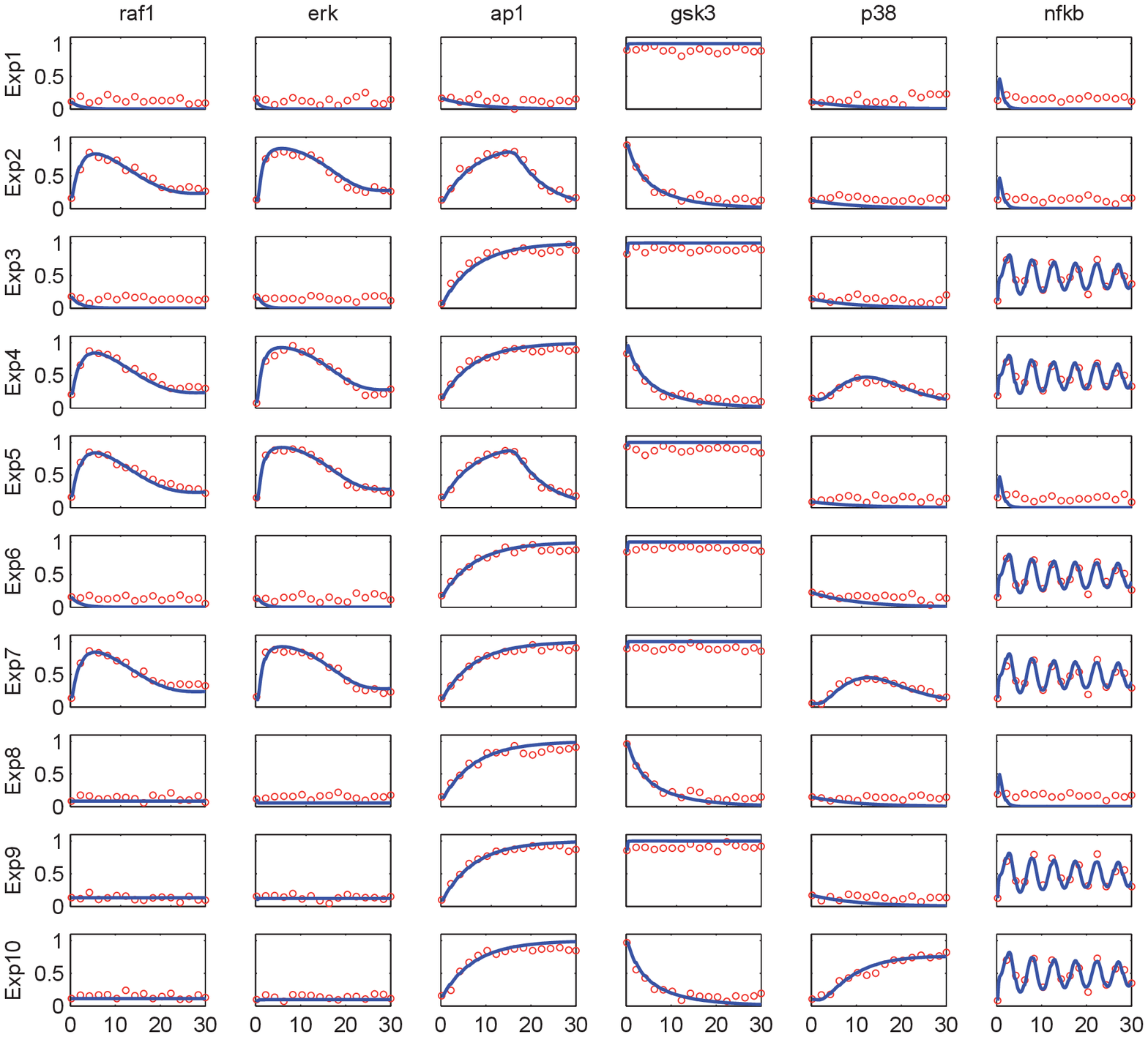} 
\caption{\textbf{Benchmark 5. Data fits: time courses.} Pseudo-experimental data (red circles) vs. optimal solution (solid blue lines) for the 6 observed states. X axis: time [minutes]. Y axis: activation level [0$\div$1]}
\label{b5_f1}
\end{figure*}
\end{center}

As an example of the fit between data and model output that is obtained after calibration, let us consider benchmark B5, which uses pseudoexperimental data corresponding to ten different experiments. Figure \ref{b5_f1} reports a good match between data and model output; notably, the algorithm manages to reproduce the oscillations in NF$\kappa$B.

\begin{center}
\begin{figure*}[ht]
\includegraphics[width=12cm]{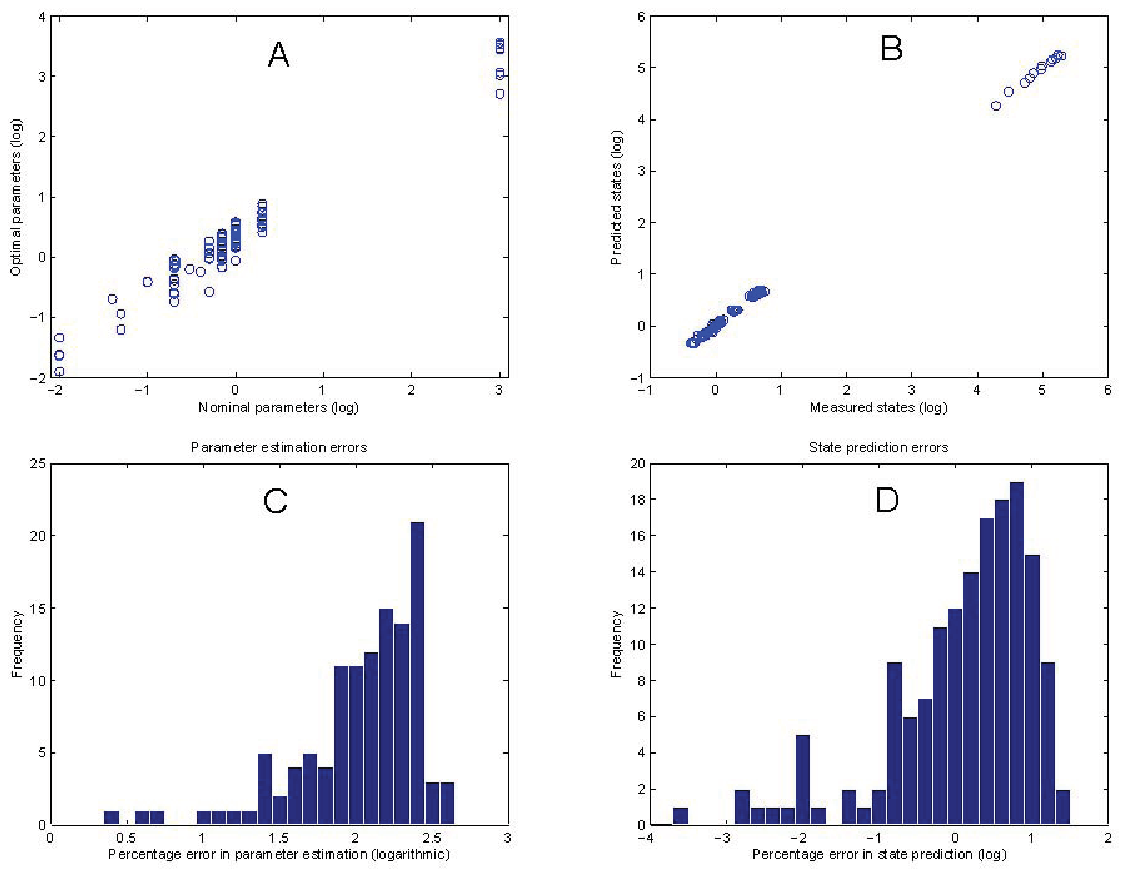} 
\caption{\textbf{Benchmark 4, typical parameter estimation results.} Left side: difference between the nominal parameter vector and the optimal solution. Top left: absolute values. Bottom left: histogram of the differences in \%. Right side: data fits (states). Top right: pseudo-experimental vs. simulated data (optimal solution). Bottom right: histogram of the state errors in \%.}
\label{b4_f4f5}
\end{figure*}
\end{center}

The fit can also be represented with histograms of the residuals, which show the distribution of the errors in the state variables. This kind of plot can also be used for showing the errors in the recovered parameters when compared to the nominal (true) values. An alternative way of visualizing this relation is by plotting the predicted states (or parameter) values as a function of the true values. This results in a diagonal--like plot; the larger the deviations from the diagonal, the larger the prediction errors. When there are identifiability issues, the fit is typically better for the states than for the parameters, because a good fit to the data does not necessarily ensure that the correct parameters have been recovered. Examples of these plots are shown in Figure \ref{b4_f4f5}, which shows the fits obtained for benchmark B4.

\section*{Conclusions}

To address the current lack of ready-to-run benchmarks for large-scale dynamic models in systems biology, we have presented here a collection of six parameter estimation problems. They cover the most common types, including metabolism, transcription, signal transduction, and development. The benchmarks are made available in a number of formats. As a common denominator, all of the models have been implemented in Matlab and C. When possible (i.e. for benchmarks B1--B5), model descriptions are also given in SBML. Ready-to-run implementations of all the benchmarks are provided in Matlab format (both with and without the AMIGO toolbox) and in COPASI (for benchmarks B1--B4). With these files it is straightforward to reproduce the results reported here. 

More importantly, the benchmark files can be easily adapted to test new parameter estimation methods for which a Matlab, C, or COPASI implementation is available. The performance of an existing or newly developed method can be evaluated by comparing its results with those reported here, as well as with those obtained by other methods. To this end, we have provided guidelines for comparing the performance of different optimizers. The problems defined here may also be used for educational purposes, running them as examples in classes or using them as assignments.

Finally, it should be noted that the utility of this collection goes beyond parameter estimation: the models provided here can also be used for benchmarking methods for optimal experimental design, identifiability analysis, sensitivity analysis, model reduction, and in the case of metabolic models also for metabolic engineering purposes. 

\section*{Competing interests}
  The authors declare that they have no competing interests.

\section*{Author's contributions}
JRB, PM, and JJ conceived of the study. 
JRB and AFV coordinated the study.
KS, SB, JS, DCS, AC, JSR, KM, PM, and JJ contributed with models.
AFV, DH, KS, and DCS formatted the models and tested the implementations.
AFV and DH carried out the computations.
AFV, EBC, and JRB analyzed the results.
AFV and JRB drafted the manuscript.
KS, SB, AC, JSR, KM, EBC, and JJ helped to draft the manuscript.
All authors read and approved the final manuscript.

\section*{Supplementary Information}
\small{\url{http://www.iim.csic.es/~gingproc/biopredynbench/}}

\section*{Acknowledgments}

This work was supported by the EU project ``BioPreDyn'' (EC FP7-KBBE-2011-5, grant number 289434). We would like to thank Attila G\'abor for reading the manuscript, providing critical comments, and finding bugs in the codes; David Rodr\'iguez Penas for helping in debugging the codes; and Thomas Cokelaer for providing the SBML qual file of model B5.


\end{document}